\documentstyle[aps,epsf]{revtex}  

\begin{document}        

\baselineskip 14pt
\title{The Future of Cherenkov Astronomy}
\author{Michael Catanese\footnote{This research is supported by the Department of Energy and NASA.}}
\address{Iowa State University}
%
\maketitle              

\begin{abstract}        
In the last ten years, the field of Cherenkov astronomy has become an
important contributor to high energy astrophysics with the detection
of eight objects at energies above 300\,GeV.  These observations have
advanced our understanding of active galactic nuclei, supernova
remnants, the extragalactic background light and cosmic-ray
acceleration and production.  Several efforts are now underway to
develop new Cherenkov telescopes which will cover a wider range of
energies (10 GeV - 50 TeV), improve the flux sensitivity by at least
an order of magnitude and provide more accurate measures of particle
energy and direction.  I describe some of the new Cherenkov telescopes
and discuss their potential to improve our understanding of
astrophysics and fundamental physics.
\end{abstract}   	

\section{Introduction}               

Cherenkov telescopes indirectly detect $\gamma$-rays by observing the
flashes of Cherenkov light emitted by particle cascades initiated when
the $\gamma$-rays interact with nuclei in the atmosphere.  These
telescopes have effective areas of 10,000\,m$^2$ to 100,000\,m$^2$,
making them efficient at detecting very short time-scale variations.
Telescopes with pixellated cameras that ``image'' the shower utilize
the differences in the Cherenkov images from $\gamma$-ray and
cosmic-ray primaries to reject the dominant cosmic-ray background with
$>$99.7\% efficiency.  These telescopes are used singly or as arrays
that stereoscopically image the Cherenkov flash and they currently
detect 250\,GeV to 20\,TeV $\gamma$-rays.  Primary particle directions
are reconstructed with accuracies of about 0.15$^\circ$ and 0.1$^\circ$
and the energy resolution is about 35\% (RMS) and 20\% for current
single telescopes and arrays, respectively.

The first clear detection of a $\gamma$-ray source by a Cherenkov
telescope was the Crab Nebula by the Whipple collaboration in 1989
\cite{Weekes89}.  At present, seven other objects have been detected
with high statistical significance: one shell-type supernova remnant
(SNR), two pulsar-powered nebulae, and four active galactic nuclei
(AGN).  Thorough reviews of the current status of the field of very
high energy (VHE, E$\ge$100\,GeV) astrophysics can be found elsewhere
\cite{vheReviews}.

\begin{figure}
\begin{minipage}[b]{3.25in}
\centerline{\epsfysize 3.5 truein \epsfbox{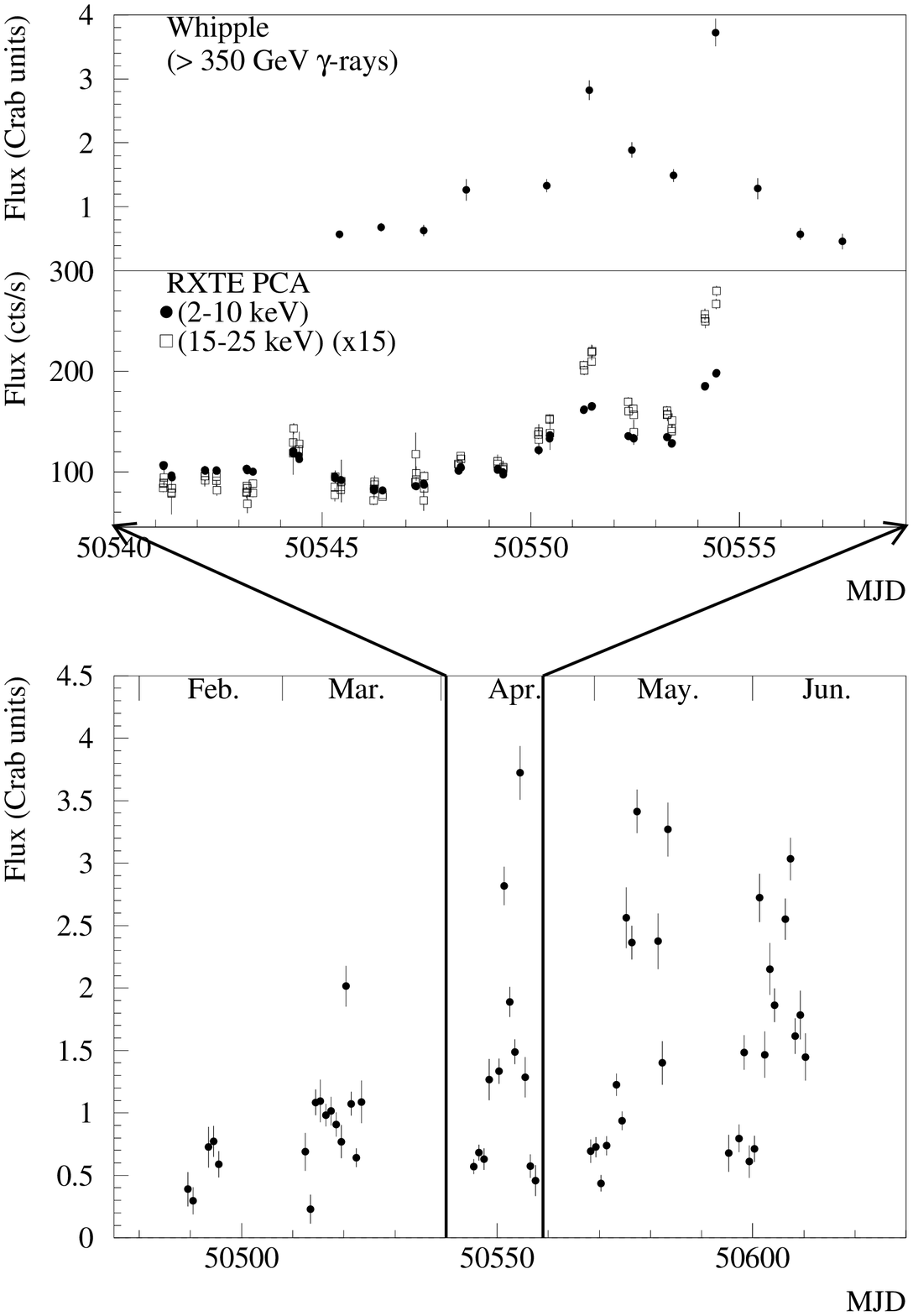}}
\caption[]{
\label{m5multi}
\small {\it Upper panel}: Observations of Mrk 501 in VHE $\gamma$-rays
(top) and X-rays (bottom) taken during 1997 April 2-20 (MJD
50540-50559).  Figure adapted from \cite{Catanese99}.  {\it Lower
panel}: Average daily $\gamma$-ray rates observed with Whipple for Mrk
501 in 1997.  Figure from \cite{Quinn99}.}
\end{minipage}
\hfill
\begin{minipage}[b]{3.25in}
\centerline{\epsfysize 2.9 truein \epsfbox{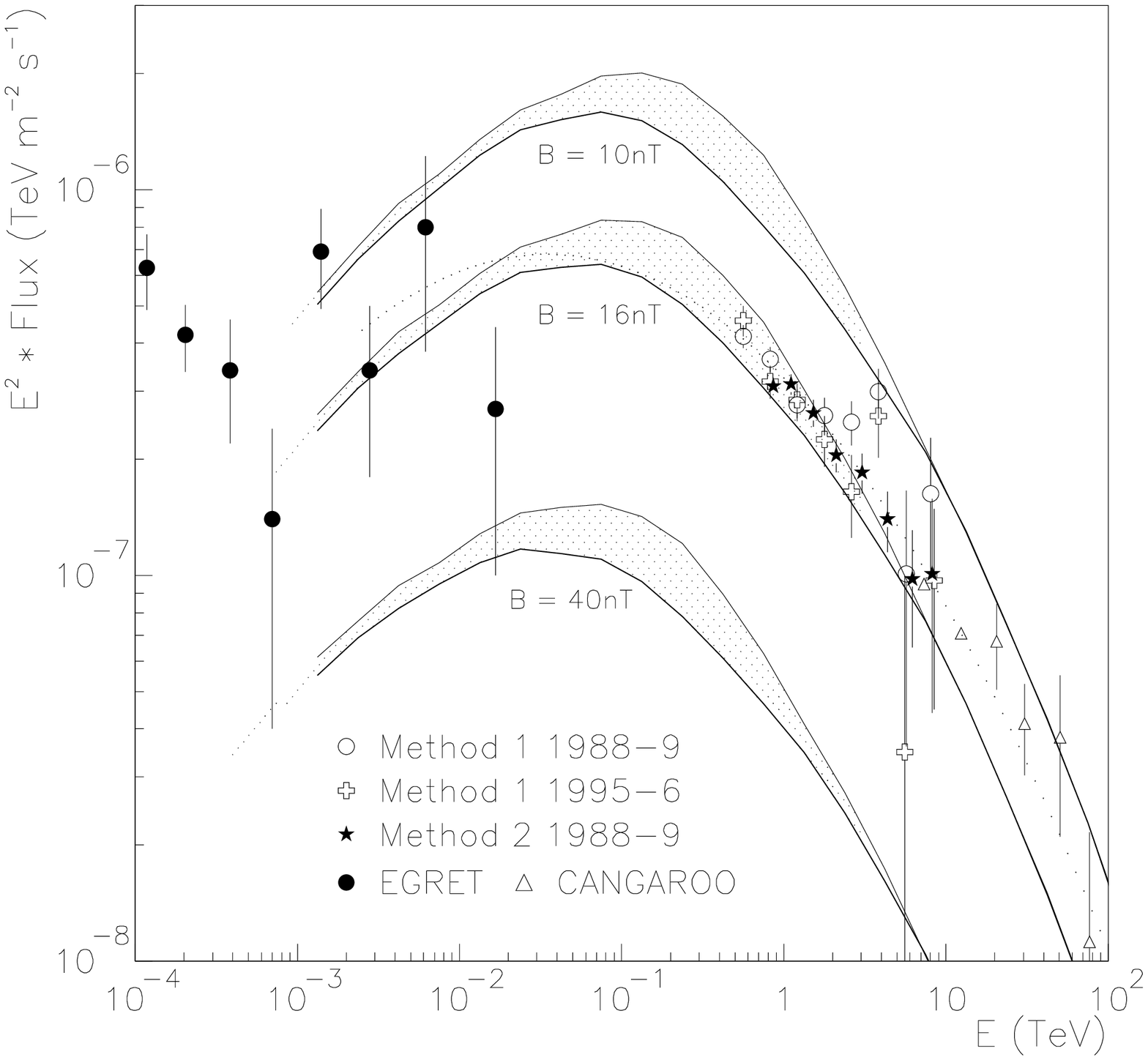}}
\vspace*{0.1in}
\caption{Crab spectrum from EGRET and VHE observations showing inverse
Compton model fits for various values of the nebular magnetic field.  Figure
from \protect\cite{Hillas98}.
\label{crab-fig}
}
\end{minipage}
\end{figure}

The observations of AGN reveal extremely large amplitude and rapid
flux variations \cite{Quinn99,Gaidos96} which correlate with
variations at longer wavelengths \cite{Macomb95} (see
Figure~\ref{m5multi}).  These provide estimates of the magnetic field
in the AGN jets and the amount of relativistic Doppler boosting of the
emission and are most easily explained if the $\gamma$-rays are
produced through inverse Compton scattering of low energy photons and
electrons.  However, the energy spectra of the AGN extend to
$>$10\,TeV \cite{VHEspectra} which is more easily explained by proton
models because the electron inverse Compton process becomes
inefficient above a few TeV.  In travelling to Earth, the TeV
$\gamma$-rays emitted by AGN are attenuated by pair-production with
optical/IR photons \cite{IRrefs}.  While this eliminates $\gamma$-rays
from very distant sources, the effect on the TeV spectra from nearby
AGN can be used to estimate the density of the extragalactic
background light (EBL) \cite{Stecker92}.  With the spectra from
Mrk\,421 and Mrk\,501, upper limits on the IR background are already,
at some wavelengths, more than 10 times better than those achieved
with direct measurements \cite{Biller98}.  VHE $\gamma$-ray
measurements of the spectrum of the Crab Nebula are consistent with
the emission being produced by inverse-Compton scattering of electrons
and photons in the synchrotron nebula (see Figure~\ref{crab-fig}) and
provide estimates of the nebular magnetic field \cite{Hillas98}.
Similarly, TeV $\gamma$-rays from the shell-type SNR, SN\,1006
\cite{Tanimori98}, provide estimates of the magnetic field and the
acceleration time of the electrons in the SNR, both previously unknown
variables in modelling the emission from this object.

Despite these exciting results, the current generation of Cherenkov
telescopes only scratches the surface of the science to which the
field can contribute.  The fact that EGRET detected over 250 objects
\cite{Hartman99} above 100\,MeV while only eight objects have been
detected above 300\,GeV indicates that much can be gained by lowering
the energy range covered by Cherenkov telescopes.  New instruments
also need to improve flux sensitivity and estimates of the
$\gamma$-ray energy and direction in order to detect more sources and
better test emission models.  Here I discuss how proposed Cherenkov
telescopes will accomplish these goals and what we hope to learn from
the data they will collect.

\section{New Cherenkov telescope projects}


\subsection{Imaging telescopes}

Proposed imaging arrays have good sensitivity from 50\,GeV to 50\,TeV.
The energy threshold is lowered by increasing the mirror area and
using a multiple telescope trigger to eliminate the background
triggers from local penetrating muons and fluctuations of the night
sky background light.  Also, because arrays measure a shower in
several telescopes, less light need be recorded in individual
telescopes to reconstruct the shower - further reducing the achievable
energy threshold.  With multiple images of the shower, its geometry
and development is better characterized, improving the angular
resolution, the ability to identify $\gamma$-ray induced showers, and
determination of the primary $\gamma$-ray energy.

\begin{figure}	
\begin{minipage}[t]{3.25in}
\centerline{\epsfysize 3.1 truein \epsfbox{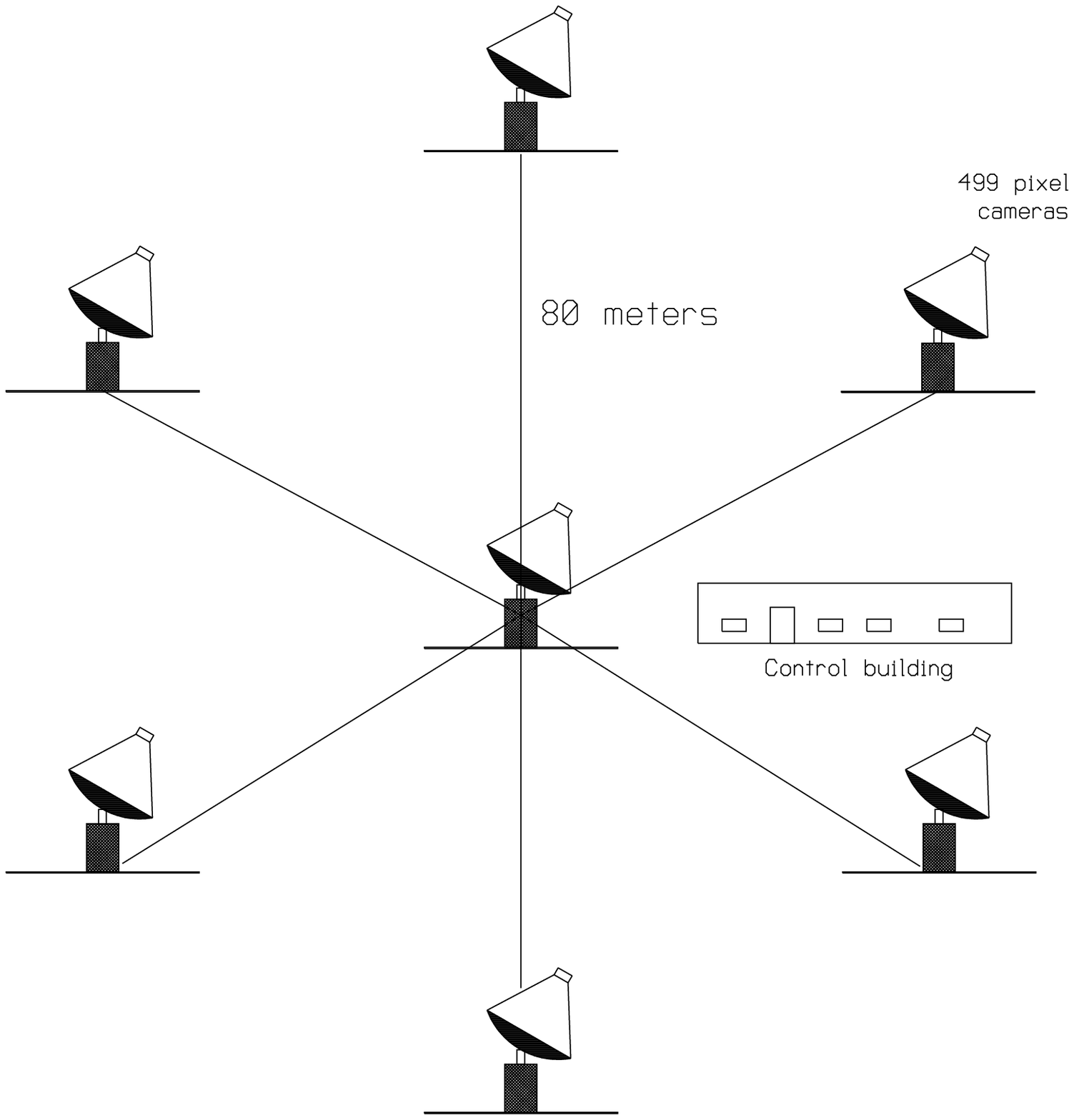}}   
\caption[]{
\label{veritas}
\small Conceptual arrangement of the telescopes for VERITAS.}
\end{minipage}
\hfill
\begin{minipage}[t]{3.25in}
\centerline{\epsfxsize 3.1 truein \epsfbox{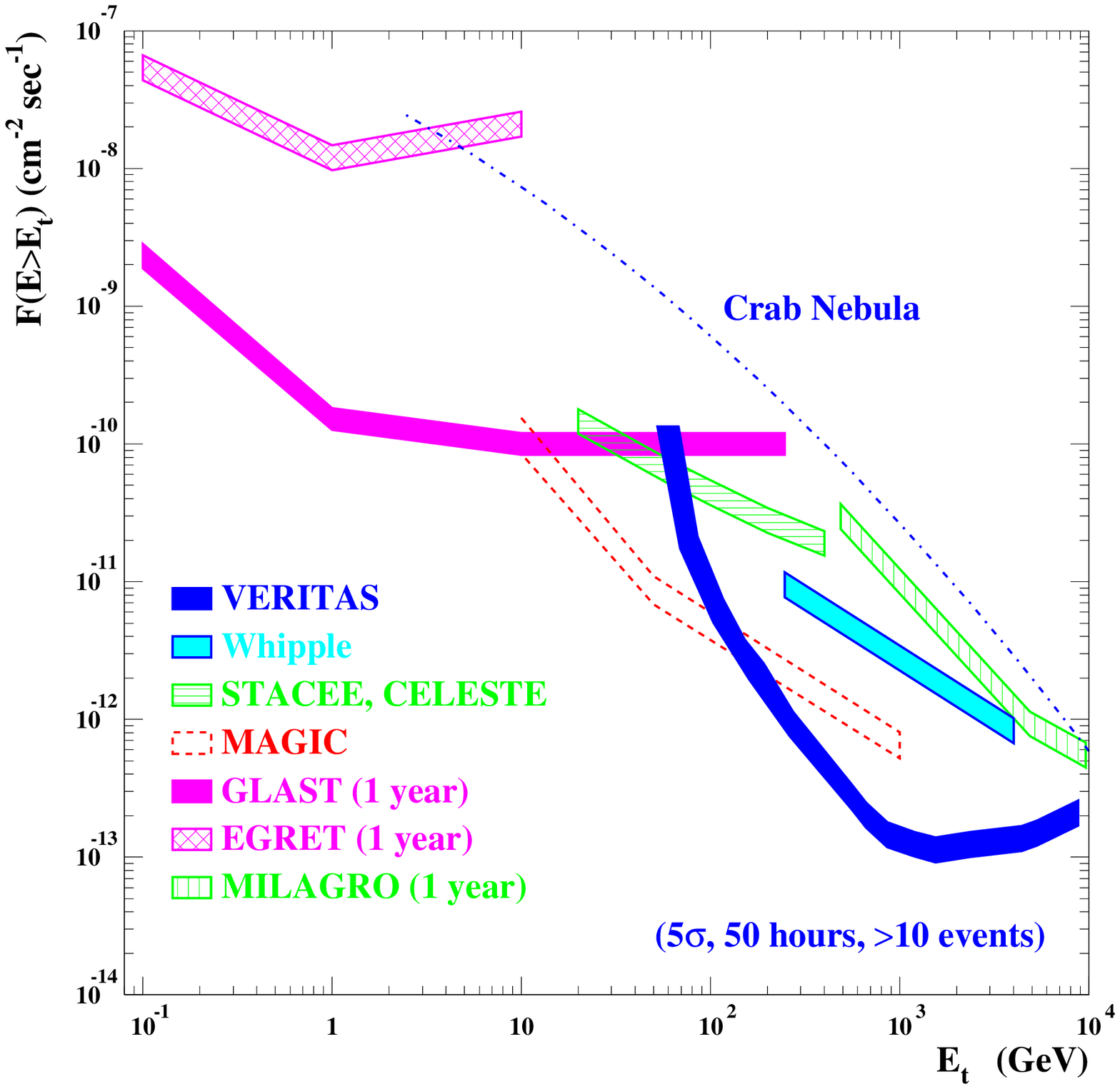}}
\caption{Point source sensitivities of several
existing and proposed $\gamma$-ray telescopes.  The sensitivities of
the pointed instruments (VERITAS, Whipple, STACEE/CELESTE, and MAGIC)
are for 50 hours of observations while the wide field instruments
(EGRET, GLAST \protect\cite{Gehrels99}, and MILAGRO
\protect\cite{Sinnis95}) are for 1 year of observations.
\label{gammasens}
}
\end{minipage}
\end{figure}

The Very Energetic Radiation Imaging Telescope Array System (VERITAS)
\cite{VERITAS} is one such proposed array of seven 10\,m telescopes
(Figure~\ref{veritas}) to be located at the base of Mt. Hopkins in the
Whipple Observatory in Arizona.  Six of the telescopes will be
arranged at the corners of a hexagon with 80\,m sides and the seventh
telescope will sit at the center.  Each telescope will have an imaging
camera of 499 photomultiplier tubes (PMTs) viewing a 3.5$^\circ$
diameter area of the sky.  An energy threshold of 75\,GeV will be
achieved and the sensitivity will be approximately 20 times better
than the current Whipple telescope.  VERITAS will have an angular
resolution of 0.09$^\circ$ at 100\,GeV which improves to 0.03$^\circ$
at 1\,TeV and its RMS energy resolution will be $<$15\%.  The High
Energy Stereoscopic System (HESS) \cite{HESS} is a proposed array with
similar performance to VERITAS, planned for operation in the southern
hemisphere, likely Namibia.  HESS could eventually consist of sixteen
10\,m diameter telescopes on a square grid with 100\,m spacing.

The Major Atmospheric Gamma Imaging Cherenkov (MAGIC) telescope
\cite{MAGIC} attempts to maximize the performance of a single imaging
telescope.  The baseline proposal for MAGIC is a 17\,m diameter
mirror, equipped with a camera of approximately 530 pixels viewing a
3.6$^\circ$ diameter field of view.  If high quantum efficiency
($\sim$45\%) hybrid PMTs become economically viable, MAGIC is
predicted to achieve an energy threshold of 30\,GeV.

\subsection{Solar arrays}

Heliostat arrays have mirror areas of several thousand square meters,
so they can efficiently detect 20 -- 300\,GeV $\gamma$-ray induced
Cherenkov flashes.  Secondary mirrors at a central tower focus the
light from individual heliostats onto PMTs (each PMT views one
heliostat) to sample the Cherenkov wavefront rather than image its
development.  Cosmic-ray background rejection is achieved by measuring
the lateral distribution of the Cherenkov light.  Two groups (CELESTE
\cite{Quebert95} and STACEE \cite{Chantell98}) have begun operation of
prototypes of these solar arrays.  During 1999, they should become
fully operational and achieve an energy threshold of $\sim$50\,GeV.

\vspace*{0.2in}

The sensitivities to point sources of the different types of telescope
operating in this energy range are shown in Figure~\ref{gammasens}.
Clearly, imaging arrays will have the greatest sensitivity in their
energy range, but the solar arrays and MAGIC can achieve lower
energy thresholds.

\section{Scientific Motivation}


\subsection{Extragalactic astrophysics}

\subsubsection{Active galactic nuclei}

\begin{figure}	
\centerline{\epsfysize 3.5 truein \epsfbox{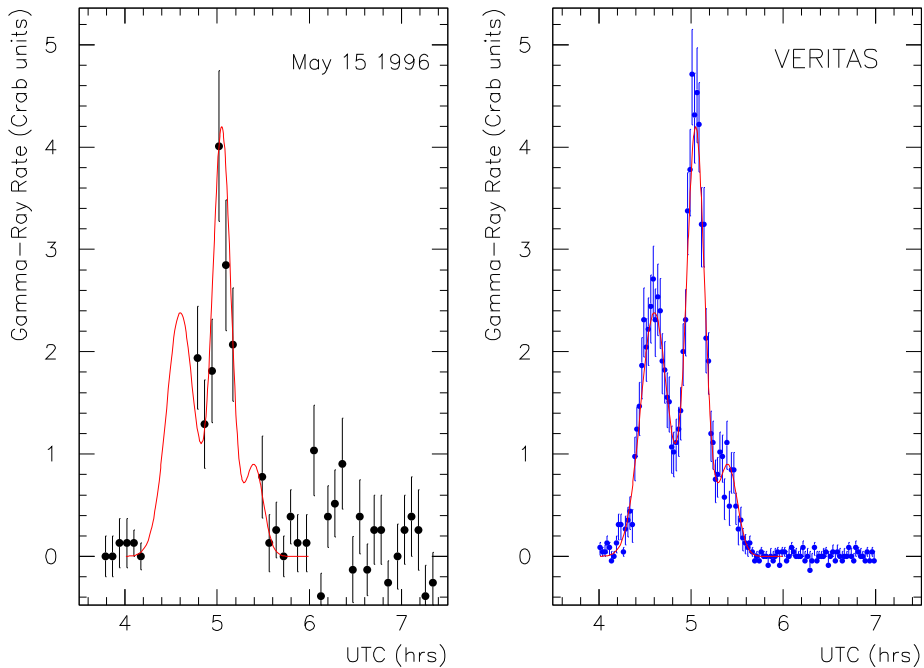}}   
\caption[]{
\label{m4hourly}
\small {\it Left}: Whipple observations of a flare from Mrk 421
on 1996 May 15 \protect\cite{Gaidos96}.  The dashed curve
is a possible intrinsic flux variation which is consistent with the
VHE data. {\it Right}: Simulated response of VERITAS to the flare
above 200 GeV.}
\end{figure}

Outstanding questions about AGN include the particle which dominates
the production of $\gamma$-rays (protons or electrons), the
mechanism by which $\gamma$-rays are produced, and the
acceleration mechanism for the particles.  Variability studies are
important to understanding the physics of the central source of AGN
because the core regions cannot be resolved with existing
interferometers.  The large effective area of Cherenkov telescopes
enables accurate measurements of extremely short variations in the
$\gamma$-ray flux as indicated in Figure~\ref{m4hourly}.  The left
part of the figure shows Whipple observations of the fastest flare
ever recorded at $\gamma$-ray energies \cite{Gaidos96}.  While the
flare is clearly detected, the structure of the flare is not resolved.
The dashed curve is a hypothetical flux variation which matches the
Whipple data.  The right part of the figure shows a simulation of how
an imaging array would clearly resolve all features of the flare.

Because blazars are extremely variable at all wavelengths, the best
way to understand the physical processes at work in them is to conduct
detailed observations spanning as wide an energy range as possible.
The new Cherenkov instruments and space-based telescopes will make
measurements spanning 6 orders of magnitude in $\gamma$-ray energies.
In addition, the arrays of telescopes will have significantly improved
energy resolution to better measure the AGN spectra which is crucial
to understanding the emission and flaring mechanisms.

The new Cherenkov telescopes should also significantly increase the
number of sources detected at VHE energies.  A lower energy threshold
will permit viewing objects further from Earth (the optical depth for
pair production with low energy photons decreases rapidly with
decreasing energy) and those objects which have spectral cut-offs
below the sensitive range of existing telescopes (e.g., EGRET
sources).  The improved flux sensitivity of the imaging arrays will
permit the detection of more of the AGN already detected with the
Cherenkov telescopes.  Measurements of the ends of the spectra for a
wide range of AGN types at different redshifts can help determine what
particles produce the $\gamma$-ray emission and refine or eliminate
unification models of blazar-type AGN \cite{Unifieds}.

\subsubsection{Infrared background radiation}

The current limits on the IR density derived from measurements of the
TeV spectra of AGN are approximately 5 to 10 times higher than
predicted from galaxy evolution \cite{IRmodels}.  However, they place
substantial restrictions on several proposed particle physics and
cosmological models which would contribute to the IR background
\cite{Biller98}.  The new Cherenkov telescopes should substantially
improve these limits to the EBL.  With a large ensemble of sources,
the energy resolution of the imaging telescope arrays may resolve the
intrinsic spectra of the AGN from the external absorption features so
that it may even be possible to detect the EBL itself.  Because the
EBL is predominantly the result of galaxy formation, these
measurements will add to our understanding of that process as well.

\subsubsection{Gamma-ray bursts}

X-ray and optical afterglows confirm that $\gamma$-ray bursts are
extragalactic but the sources and mechanism for producing the
$\gamma$-ray bursts remain unknown.  The delayed GeV photons from
$\gamma$-ray bursts \cite{Hurley94} demonstrate that high energy
$\gamma$-rays play an important role in $\gamma$-ray bursts that can
be pursued with rapid follow-up observations.  With low energy
thresholds, new Cherenkov telescopes will be able to see bursts out to
z$\sim$1 or more.  Because of the difficulty in producing VHE
$\gamma$-rays and in getting them out of the region where the burst
originates, the detection of a VHE component would place stringent
limits on the viable models for $\gamma$-ray bursts.  Attenuation from
interaction with the EBL can also provide an independent distance
estimate if optical follow-up observations do not reveal spectral
lines.

\subsection{Galactic astrophysics}

\subsubsection{Shell-type supernova remnants and cosmic rays}


SNRs are widely believed to be the sources of hadronic cosmic rays up
to energies of approximately $Z\times 10^{14}$\,eV, where $Z$ is the
nuclear charge of the particle.  The existence of energetic {\it
electrons} in SNRs is well-known from observations of synchrotron
emission at radio and X-ray wavelengths and TeV
$\gamma$-rays from SN\,1006 \cite{Tanimori98}, most likely generated
by electrons through inverse Compton scattering.  However, a clear
indication for the acceleration of hadronic particles in SNR is
lacking.  The evidence for such particles would be a characteristic
spectrum of $\gamma$-rays produced mostly via $\pi^0$ decay subsequent
to nuclear interactions in the SNR.  While EGRET has detected signals
from several regions of the sky that are consistent with the positions
of shell-type SNRs \cite{EGRETsnrs}, upper limits from the Whipple
collaboration at E$>$300\,GeV are well below the extension of the
EGRET spectra \cite{Buckley98}.

\begin{figure}
\begin{minipage}[b]{3.25in}
\centerline{\epsfxsize 3.2 truein \epsfbox{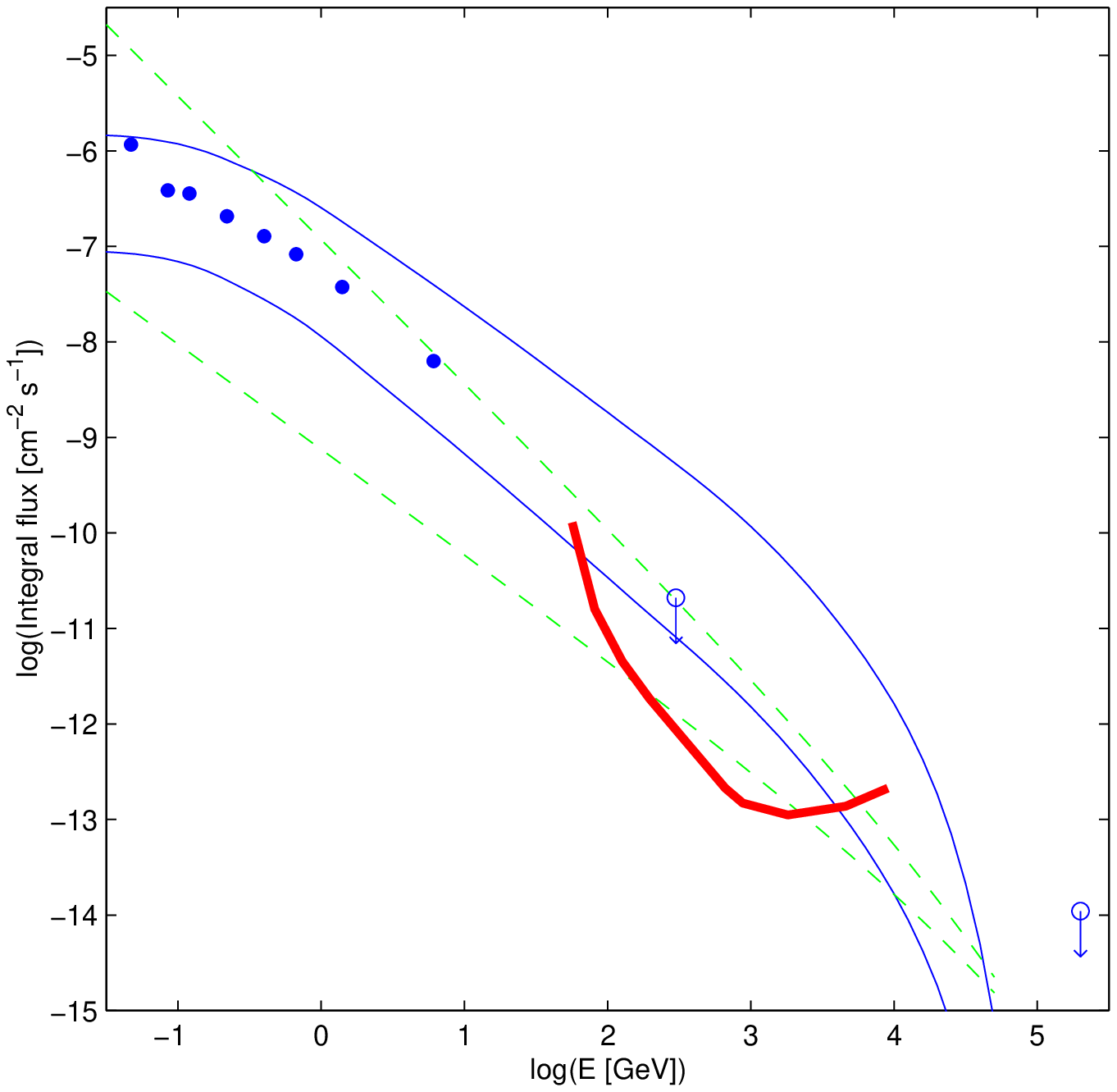}}
\caption{Predicted $\gamma$-ray spectra in the shell-type SNR IC\,443.
The solid lines depict a range of spectra from hadronic interactions
(adapted from \protect\cite{Buckley98}).  The dashed lines depict
inverse-Compton spectra for a range of parameters allowed by X-ray
data.  EGRET data (filled circles) and upper limits (open circles)
from Whipple and Cygnus are shown.  The predicted
sensitivity of VERITAS for a 50 hour observation is indicated by the
thick curve.
\label{main-snrlims-fig}
}
\end{minipage}
\hfill
\begin{minipage}[b]{3.25in}
\centerline{\epsfxsize 3.2 truein \epsfbox{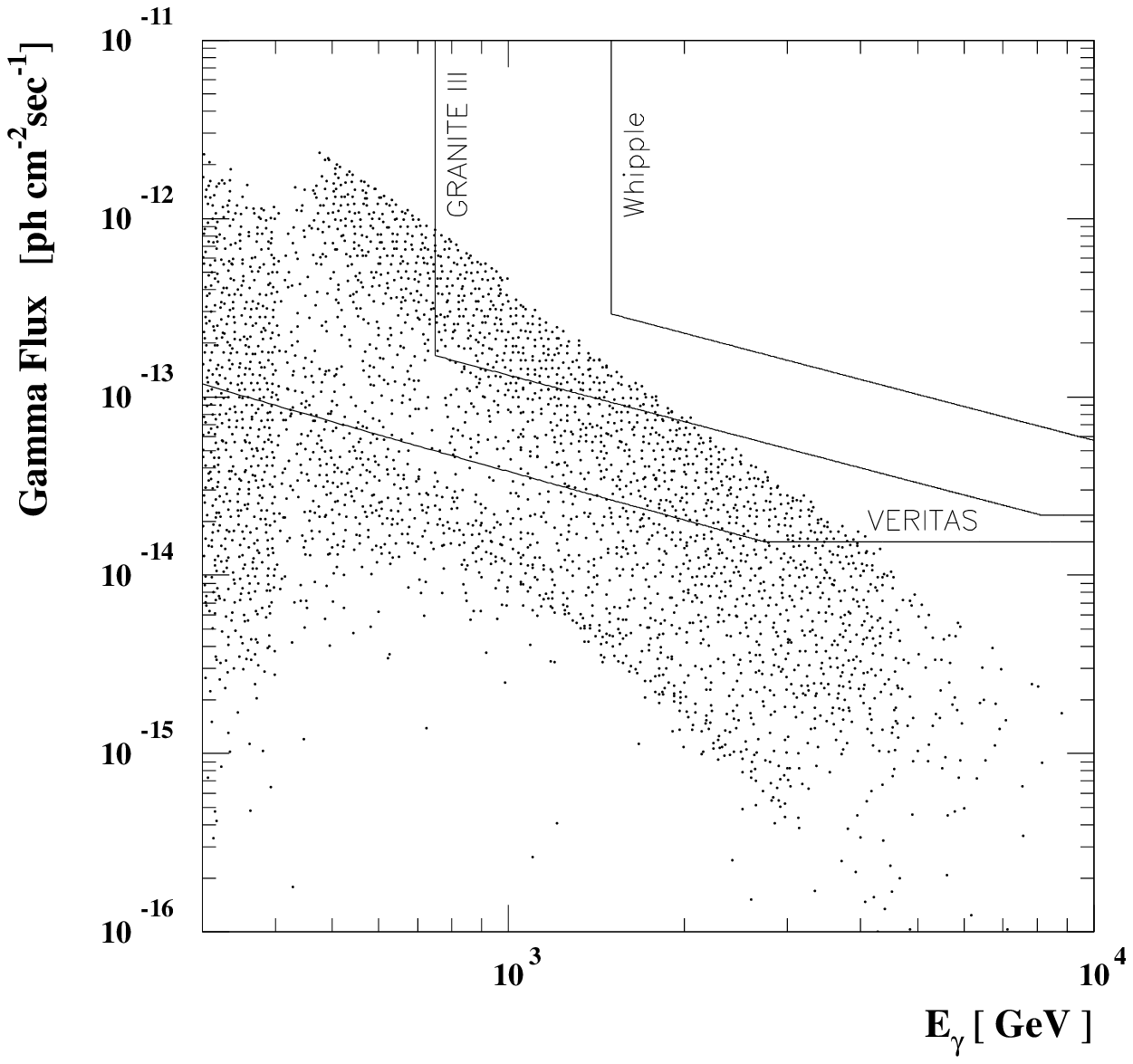}}
\caption{Neutralino annihilation rate from the Galactic center.
Shown are the sensitivity curves for Whipple, the Granite III upgrade
of the Whipple telescope, and VERITAS.  Each point represents a
particular choice of supersymmetry model parameters in the
experimentally allowed range.  Figure from \protect\cite{Bergstrom98}.
\label{main-neut-fig}
}
\end{minipage}
\end{figure}

As shown in Figure~\ref{main-snrlims-fig}, there are predictions for
strong $\gamma$-ray emission from shell-type SNRs by hadron {\it and}
electron interactions.  Model fits to EGRET and Whipple data
\cite{Gaisser98} indicate that if the emission detected by EGRET is
from the SNR, inverse Compton and bremsstrahlung scattering of
electrons contribute to the flux and the hadronic spectrum is steeper
than the $E^{-2.1}$ expected from direct cosmic-ray measurements.  The
new Cherenkov telescopes, particularly the imaging arrays, and GLAST
will provide excellent sensitivity and energy reconstruction for
resolving the various emission components in these objects.  In
addition, the imaging arrays will provide detailed mapping of the
emission regions in the SNRs.  For a typical SNR luminosity and
angular extent, an imaging array should be able to detect
approximately 20 objects within 4\,kpc of Earth according to one
popular model of $\gamma$-ray production by hadronic interactions
\cite{Drury94}, permitting investigation of which characteristics in
SNR are necessary for particle acceleration.

\subsubsection{Compact Galactic Objects}
\label{compact-gal-sect}

VHE emission from the Crab, PSR\,1706-44 and Vela suggest that they
may be the most prominent members of a large galactic population of
sources.  An accurate VHE spectrum is crucial to understanding the
production mechanism of $\gamma$-rays from these pulsar-powered
nebulae.  The new imaging arrays should be sensitive to Crab-like
objects anywhere within the Galaxy.  The energy resolution of the
arrays and the broad energy coverage available by combining the data
with GLAST measurements will significantly improve tests of
$\gamma$-ray emission models.  Finally, the imaging arrays may even be
able to resolve the VHE emission region of nearby objects like the
Crab Nebula.

VHE $\gamma$-rays produced near a pulsar will pair produce with the
intense magnetic fields there, leading to a sharp spectral cut-off.
Thus, VHE observations constrain the location of the pulsar particle
acceleration region.  The high energy emission of the six pulsars
detected at EGRET energies \cite{Hartman99} is already seriously
constrained by the VHE upper limits \cite{vheReviews}.  The energy
threshold of the new telescopes should permit the detection of these
bright GeV sources.

Of the EGRET sources, 170 have no known counterpart at longer
wavelengths \cite{Hartman99}, mostly due to their positional
uncertainty.  With their sensitivity and energy threshold, Cherenkov
telescopes should detect many of these objects and
source locations from imaging arrays could lead to identifications
with objects at longer wavelengths.

A survey is an efficient means of observing a large sample of sources
and the only way to efficiently detect new types of sources.  Imaging
arrays will be able to survey the sky in the 100\,GeV -- 10\,TeV
energy range.  An 80-night survey of the Galactic plane region
$0^\circ < l < 85^\circ$ with VERITAS will be sensitive to fluxes down
to $\sim$0.02 Crab above 300\,GeV and encompass more than 40 potential
VHE sources, and so should significantly increase the VHE catalog.

\subsection{Fundamental Physics}

\subsubsection{Neutralino annihilation in the Galactic center}

Current astrophysical data indicate the need for a cold dark matter
component with $\Omega \approx 0.3$.  A good candidate for this
component is the neutralino, the lightest stable supersymmetric
particle.  If neutralinos do comprise the dark matter and are
concentrated near the center of our galaxy, their direct annihilation
to $\gamma$-rays should produce a monoenergetic annihilation line with
mean energy equal to the neutralino mass.  Cosmological constraints
and limits from accelerator experiments restrict the neutralino mass
to the range 30\,GeV - 3\,TeV.  Thus, the new Cherenkov telescopes and
GLAST together will allow a sensitive search over the entire allowed
neutralino mass range.  Recent estimates of the annihilation line flux
for neutralinos at the galactic center \cite{Bergstrom98} predict a
$\gamma$-ray signal which may be of sufficient intensity to be
detected with an imaging array (Figure~\ref{main-neut-fig}) and GLAST.

\subsubsection{Quantum gravity}

Quantum gravity can manifest itself as an effective energy-dependence
to the velocity of light in vacuum caused by propagation through a
gravitational medium containing quantum fluctuations.  In some
formulations \cite{Amelino98}, this time dispersion can have a
first-order dependence on photon energy:
\begin{equation}
\Delta t \simeq \xi \frac{E}{E_{QG}} \frac{L}{c}
\label{main-qg-eq}
\end{equation}
where $\Delta t$ is the time delay relative to the energy-independent
speed of light, $c$; $\xi$ is a model-dependent factor of order 1; $E$
is the energy of the observed radiation; $E_{QG}$ is the energy scale
at which quantum gravity couples to electromagnetic radiation; and $L$
is the distance over which the radiation has propagated.  Recent work
within the context of string theory indicates that quantum gravity may
begin to manifest itself at a much lower energy scale than the Planck
mass, perhaps as low as 10$^{16}$\,GeV \cite{Witten96}.  VHE
observations of variable emission from distant objects provide an
excellent means of searching for the effects of quantum gravity.  For
example, the Whipple Collaboration has recently used data from a rapid
TeV flare of the AGN Mrk\,421 to constrain $E_{QG}/\xi$ to be $>4
\times 10^{16}$\,GeV, the highest convincing limit determined to date
\cite{Biller99}.  This limit can be vastly improved with the new
Cherenkov telescopes because they will be more sensitive to short
time-scale variability and able to detect more distant objects.  In
addition to AGN flares, $\gamma$-ray bursts and pulsed emission from
Galactic sources may provide avenues for investigating the effects of
quantum gravity.


\begin{references}  

\bibitem{Weekes89} T. C. Weekes {\it et al.}, Astrophys. J. {\bf 342},
379 (1989).







\bibitem{vheReviews} T. C. Weekes, F. A. Aharonian, D. J. Fegan, and
T. Kifune, in {\it Proc. of the Fourth Compton Symp.}, edited by
C. D. Dermer, M. S. Strickman, and J. D. Kurfess (American Inst. of
Phys., New York, 1997), AIP Conf. Proc. 410, p. 361; R. A. Ong,
Physics Reports {\bf 305}, 95 (1998).

\bibitem{Quinn99} J. Quinn {\it et al.}, Astrophys. J. (to be published).

\bibitem{Gaidos96} J. A. Gaidos {\it et al.}, Nature {\bf 383}, 319 (1996).

\bibitem{Macomb95} J. H. Buckley {\it et al.}, Astrophys. J. {\bf 472}, 
L9 (1996); M. Catanese {\it et al.}, Astrophys. J. {\bf 487}, L143
(1997).

\bibitem{VHEspectra} F. A. Aharonian {\it et al.}, Astron. and Astrophys.
{\bf 327}, L5 (1997); F. Krennrich {\it et al.}, Astrophys. J. {\bf 511}, 149 (1999).

\bibitem{Catanese99} M. Catanese, in {\it BL Lac Phenomenon}, proceedings of
the International Conference, Turku, Finland (to be published).

\bibitem{IRrefs} R. P. Gould and G. P. Schr\'eder, Phys. Rev.
{\bf 155}, 1408 (1967).

\bibitem{Stecker92} F. W. Stecker, O. C. de Jager, and M. H. Salamon,
Astrophys. J. {\bf 390}, L49 (1992).

\bibitem{Biller98} S. D. Biller {\it et al.}, Phys. Rev. Lett. 
{\bf 80}, 2992 (1998).

\bibitem{Hillas98} A. M. Hillas {\it et al.}, Astrophys. J. {\bf 503}, 
744 (1998).

\bibitem{Tanimori98} T. Tanimori {\it et al.}, Astrophys. J. {\bf 497}, L25
(1998).


\bibitem{Hartman99} R. C. Hartman {\it et al.}, Astrophys. J. Supp. (to be 
published).

\bibitem{VERITAS} R. Lessard, in {\it TeV Astrophysics of
Extragalactic Sources}, Astropart. Phys., edited by M. Catanese and
T. C. Weekes (to be published).

\bibitem{HESS} F. A. Aharonian {\it et al.}, High Energy Stereoscopic 
System (HESS) Letter of Intent, MPI f\"ur Kernphysik, Heidelberg (1997).

\bibitem{MAGIC} J. A. Barrio {\it et al.}, The Magic Telescope, design
study, MPI f\"ur Physik, Munich (1998).


\bibitem{Quebert95} J. Quebert {\it et al.}, in {\it Towards a 
Major Atmospheric Cherenkov Detector IV}, Padova, Italy, edited
by M. Cresti, 248 (1995).

\bibitem{Chantell98} M. C. Chantell {\it et al.}, Nucl. Instrum. and
Meth. A {\bf 408}, 468 (1998).

\bibitem{Gehrels99} N. Gehrels and P. Michelson, in {\it TeV Astrophysics
of Extragalactic Sources}, Astroparticle Physics, edited by M. Catanese
and T. C. Weekes (to be published).

\bibitem{Sinnis95} G. Sinnis {\it et al.}, Nucl. Phys. B (Proc. Suppl.)
{\bf 43}, 141 (1995).

\bibitem{Unifieds} G. Ghisellini {\it et al.}, Mon. Not. Roy. Astron.
Soc. {\bf 301}, 451 (1998); G. Georganopoulos and A. P. Marscher,
Astrophys. J. {\bf 506}, 621 (1998).

\bibitem{IRmodels} M. A. Malkan and F. W. Stecker, Astrophys. J. {\bf
496}, 13 (1998); J. R. Primack, J. S. Bullock, R. S. Somerville, and
D. MacMinn, in {\it TeV Astrophysics of Extragalactic Sources},
Astropart. Phys., edited by M. Catanese and T. C. Weekes (to be
published).

\bibitem{Hurley94} K. Hurley, Nature {\bf 372}, 652 (1994).




\bibitem{EGRETsnrs} J. A. Sturner and C. D. Dermer, Astron. and
Astrophys.  {\bf 293}, L17 (1995); J. A. Esposito {\it et al.},
Astrophys. J.  {\bf 461}, 820 (1996); T. R. Jaffe {\it et al.},
{\it ibid.} {\bf 484}, L129 (1997); R. C. Lamb and D. J. Macomb,
{\it ibid.} {\bf 488}, 872 (1997).

\bibitem{Buckley98} J. H. Buckley {\it et al.}, Astron. and Astrophys.
{\bf 329}, 639 (1998).

\bibitem{Gaisser98} T. K. Gaisser, R. J. Protheroe, and T. Stanev,
Astrophys. J. {\bf 492}, 219 (1998).

\bibitem{Drury94} L. O'C. Drury, F. A. Aharonian, and H. J. V\"olk,
Astron. and Astrophys. {\bf 287}, 959 (1994).



\bibitem{Bergstrom98} L. Bergstr\"om, P. Ullio, and J. H. Buckley,
Astropart. Phys. {\bf 9}, 137 (1998).


\bibitem{Amelino98} G. Amelino-Camelia {\it et al.}, Nature {\bf 383},
319 (1998).


\bibitem{Witten96} E. Witten, Nucl. Phys. B {\bf 471}, 135
(1996).

\bibitem{Biller99} S. Biller {\it et al.}, Phys. Rev. Lett. (to be published).







\end{references}
\end{document}